\newif\ifsubmode
\submodefalse

\newif\ifprintfig
\printfigtrue

\ifsubmode
  \documentstyle[12pt,aasms4,epsf]{article}
  \received{}
  \accepted{}
  \journalid{}{}
  \articleid{}{}
\else
  \documentstyle[11pt,aaspp4,epsf]{article}
  \slugcomment{submitted for publication in ApJ}
\fi

\lefthead{van den Bosch, F.C.}
\righthead{The origin of the Hubble sequence}

\newcommand{\etal}{{et al.~}}

\newcommand{\lta}{\lesssim}
\newcommand{\gta}{\gtrsim}

\newcommand{\kms}{\>{\rm km}\,{\rm s}^{-1}}

\newcommand{\Msun}{\>{\rm M_{\odot}}}
\newcommand{\Lsun}{\>{\rm L_{\odot}}}

\begin{document}

\title{The formation of disk-bulge-halo systems\\ and the origin of the
Hubble sequence}
       
\author{Frank C. van den Bosch\altaffilmark{1,2}} 
\affil{Department of Astronomy, University of Washington, Box 351580, 
       Seattle, WA 98195, USA}


\altaffiltext{1}{Hubble Fellow}
\altaffiltext{2}{vdbosch@hermes.astro.washington.edu}


\ifsubmode\else
\clearpage\fi


\ifsubmode\else
\baselineskip=14pt
\fi


\begin{abstract}
  We investigate the formation of disk-bulge-halo systems by including
  bulges in the Fall \& Efstathiou theory of disk formation.  This
  allows an investigation of bulge dominated disk galaxies, such as
  S0s and disky ellipticals. These latter systems, which consist of an
  elliptical spheroid with an embedded disk with a scale-length of
  typically a few hundred parsecs, seem to form a smooth sequence with
  spirals and S0s towards lower disk-to-bulge ratio. The aim of this
  paper is to examine whether spirals, S0s, and disky elllipticals all
  can be incorporated in one simple galaxy formation scenario.  We
  investigate an inside-out formation scenario in which subsequent
  layers of gas cool and form stars inside a virialized dark halo.
  The inner, low angular momentum material is assumed to form the
  bulge.  Stability arguments are used to suggest that this bulge
  formation is a self-regulating process in which the bulge grows
  until it is massive enough to allow the remaining gas to form a
  stable disk component.  We assume that the baryons that build the
  disk do not loose their specific angular momentum, and we search for
  the parameters and physical processes that determine the
  disk-to-bulge ratio, and therewith explain to a large extent the
  origin of the Hubble sequence. The spread in halo angular momenta
  coupled with a spread in the formation redshifts can explain the
  observed spread in disk properties and disk-to-bulge ratios from
  spirals to S0s.  If galaxy formation is efficient, and all available
  baryons are transformed into the disk-bulge system, cosmologies with
  $\Omega_0 \lesssim 0.3$ can be excluded, since stable spiral disks
  would not be allowed to form.  However, if we assume that the
  efficiency with which galaxies form depends on the formation
  redshift, as suggested by the small amount of scatter in the
  observed Tully-Fisher relation, and we assume that the probability
  for a certain baryon to ultimately end up in the disk or bulge is
  independent of its specific angular momentum, spirals are allowed to
  form, but only at small formation redshifts ($z \lesssim 1$).  At
  higher formation redshifts, stability arguments suggest the
  formation of systems with smaller disk-to-bulge ratios, such as S0s.
  Since density perturbations in clusters will generally collapse
  earlier than those in the field, this scenario naturally predicts a
  density-morphology relation, the amplitude of which depends on the
  baryon fraction of the Universe. Disky ellipticals are too compact
  to be incorporated in this scenario, and they thus do not form a
  continuous sequence with spirals and S0s, at least not in the sense of
  the galaxy formation scenario envisioned in this paper.  Alternative
  formation scenarios for the disky ellipticals, such as gas-rich
  mergers or an internal mass loss origin for the embedded disks, are
  much more viable.
\end{abstract}



\clearpage


\section{Introduction}

One of the most compelling puzzles in present day astronomy is the
question how galaxies formed. In particular, we need to understand the
wide variety of sizes, masses and morphologies of galaxies observed,
as well as their dynamics, ages and metallicities.  In addition, we
need to understand the origin of the scaling relations, such as the
fundamental plane relations for ellipticals and the Tully-Fisher
relation for spirals, as well as the so-called density-morphology
relation (Dressler 1980). The latter shows that the more compact
galaxies, such as ellipticals and S0s, are preferentially found in
overdense regions such as galaxy clusters.

The classical way of depicting the different morphologies of galaxies
is by means of the Hubble diagram, which reveals the gross distinction
of three sorts of galaxies: spirals, S0s, and ellipticals.  Several
studies over the past years have shown that elliptical galaxies can be
roughly divided in two subclasses: the rotation supported, low
luminosity disky ellipticals and the pressure supported, bright, boxy
ellipticals (e.g., Bender 1988; Bender \etal 1989; Capaccioli, Caon \&
Rampazzo 1990).  This dichotomy has recently been strengthened by
properties observed in their central regions (e.g., Nieto, Bender \&
Surma 1991; Jaffe \etal 1994; Ferrarese \etal 1994; Gebhardt \etal
1996; Faber \etal 1997).  Based on this dichotomy amongst elliptical
galaxies, it has been suggested that the classical Hubble diagram
should be revised (Kormendy \& Bender 1996), to use a more physical
subdivision of the class of the ellipticals, rather than the apparent
flattening.  Kormendy \& Bender proposed to use the velocity
anisotropy which is well measured by the isophote shapes (Bender 1988;
Bender \etal 1989).

The main morphological parameter that sets the classification of
galaxies in the (revised) Hubble diagram is the disk-to-bulge ratio
$D/B$.  Understanding the origin of the Hubble sequence is thus
intimately related to understanding the parameters and processes that
determine the ratio between the masses of disk and bulge.  The
diskiness of the low luminosity ellipticals is generally interpreted
as due to an embedded disk. These disks, which we term `embedded
disks' in the following, are smaller and brighter than disks in S0s
and spirals (Scorza \& Bender 1995), and it has been emphasized many
times that the disky ellipticals build a continuous sequence with S0
galaxies (Capaccioli 1987; Carter 1987; Bender 1988, 1990; van
den Bergh 1990: Capaccioli \etal 1990; Rix \& White 1990; Capaccioli
\& Caon 1992; Bender \etal 1993; Scorza \& Bender 1995, 1996; Scorza
\& van den Bosch 1998).  This continuity suggests similar formation
histories, whereby one or several parameters of the proto-galaxy vary
smoothly.

Disk dominated systems such as spirals are believed to have formed by
cooling of the baryonic matter inside a virialized dark halo.  As the
gas cools, its specific angular momentum is conserved, and the amount
of angular momentum of the dark halo thus determines the size of the
disk (Fall \& Efstathiou 1980; see Section~2 for more details on this
disk-formation scenario). The formation of bulge dominated disk
systems is far less clear.  The aim of this paper is to investigate
to what extent the disky ellipticals and S0s can be incorporated in
the Fall \& Efstathiou theory for the formation of galactic disks,
by incorporating a simple picture for the formation of the bulge.

We envision an inside-out formation scenario for the bulge.  It is
assumed that the bulge forms out of the low-angular momentum material
in the halo, which cools and tries to settle into a small, compact
disk.  Such disks are however unstable, and we assume here that this
instability, coupled with the continuous supply of new layers of
baryonic matter that cool and collapse, forms the bulge. This
inside-out bulge formation is self-regulated in that the bulge grows
until it is massive enough to allow the remaining gas to form a stable
disk component.  We do not describe the bulge formation in any detail
but merely use empirical relations of the characteristic structural
parameters of bulges and ellipticals to describe the end result of the
formation process as a realistic galaxy.  We use this simple formation
scenario to investigate the predicted disk-to-bulge ratios and disk
scale-lengths as a function of the halo angular momentum, and as a
function of formation redshift and cosmology. The main focus of this
paper is to investigate whether this inside-out formation scenario can
account for two orders of magnitude variation in disk-to-bulge ratio,
required in order to incorporate a wide variety of disk-bulge systems:
from late-type spirals with $D/B \gta 10$ to S0s ($D/B \sim 0.1$) to
disky ellipticals ($D/B \sim 0.1$).

This paper is organized as follows. In Section~2 we describe the
formation scenario, the galaxy formation efficiency, and the stability
criterion for our disk-bulge-halo models. In Section~3 we use these
models to investigate the position of different sorts of disks in the
parameter space of disk central surface brightness versus disk
scale-length.  In Section~4 we discuss to what extent S0s and disky
ellipticals can be incorporated in this formation scenario, and what
parameters may be responsible for the origin of (the major part of)
the Hubble sequence. In Section~5 we briefly discuss alternative
formation scenarios for disky ellipticals. Our results are summarized
and discussed in Section~6.

\section{The formation scenario}

A remarkably successful model for the formation of galaxies is the
White \& Rees (1978) theory, wherein galaxies form through the
hierarchical clustering of dark matter and subsequent dissipational
settling of gaseous matter within the dark halo cores.  Coupled with
the notion of angular momentum gain by tidal torques induced by nearby
proto-galaxies (Hoyle 1953; Peebles 1969), the White \& Rees theory
provides the background for a model for the formation of galactic
disks (Fall \& Efstathiou 1980; Faber 1982; Gunn 1982; Fall 1983; van
der Kruit 1987). In this model, disks form through the collapse of a
uniformly rotating proto-galaxy.  Owing to the non-dissipative
character of the dark matter, its collapse halts when the system
virializes. At regions where the baryonic density is high enough, the
baryons can cool and decouple from the dark matter. The angular
momentum of the baryons, acquired from tidal torques from nearby
proto-galaxies, prevents the collapse from proceeding all the way to
the center, and causes the baryons to settle in a rapidly rotating
disk.  This simple picture of disk formation yields realistic disk
sizes and surface brightness profiles provided that there is little
angular momentum transfer from the baryonic matter to the dark halo,
and that the dark halo has a density profile such that the resulting
galaxy has a flat rotation curve. Fall (1983) extended the original
Fall \& Efstathiou disk formation scenario by placing it in a
cosmological context, and by deriving Freeman's law (Freeman 1970) and
the Tully-Fisher relation. Recently, two papers reexamined this disk
formation model and included the effects of halo-contraction due to
the accumulation of baryonic matter in the center of the galaxy:
Dalcanton, Spergel \& Summers (1997) showed that the disk formation
model outlined above can not only explain the properties of normal,
high surface brightness (HSB) disks, but also those of the class of
low surface brightness (LSB) disks (see also Jimenez \etal 1998); Mo,
Mao \& White (1998) compared the outcome of disks in different
cosmologies, and used this to investigate the effects of disk
evolution and the predicted population of Ly$\alpha$ absorbers in QSO
spectra.  The assumptions we make are similar to those made by
Dalcanton \etal (1997) and Mo \etal (1998), but we extend upon their
work by including bulges.  We investigate how the disky ellipticals
and S0s may be fitted into this particular formation scenario, by
comparing the differences in properties of the proto-galaxies that
result in spirals, S0s, and disky ellipticals.

\subsection{Inside-out bulge formation}

The turn-around, virialization, and subsequent dissipational settling
of the baryonic matter of a proto-galaxy is an inside-out process.
First the innermost shells virialize and heat its baryonic material to
the virial temperature. Because of the relatively high density and low
angular momentum of this gas, it will rapidly cool and (try to) settle
in a very small, compact disk. However, self-gravitating disks are
violently unstable and form a bar (e.g., Hohl 1971; Ostriker \&
Peebles 1973).  Meanwhile the next shell of the proto-galaxy cools and
tries to settle into a disk structure at a radius determined by its
angular momentum.  The resulting structure in the center of the dark
halo consequently becomes more and more unstable.

Bars are efficient in transporting gas inwards (e.g., Friedli \&
Martinet 1993; Wada \& Habe 1995), and can cause vertical heating by
means of a collective bending instability (e.g., Combes \etal 1990;
Pfenniger 1984; Pfenniger \& Friedli 1991; Raha \etal 1991).  Both
these process lead ultimately to the dissolution of the bar; first the
bar takes a hotter and triaxial shape, but is later on transformed in
a spheroidal bulge component (e.g., Combes \etal 1990; Pfenniger \&
Norman 1990; Pfenniger \& Friedli 1991; Pfenniger 1993; Friedli 1994;
Martinet 1995; Norman, Sellwood \& Hasan 1996). Here we assume that
this bar instability and subsequent dissolution, coupled with the
continuous supply of new layers of baryonic matter that cool and
collapse, forms the bulge. The numerical simulations performed by the
numerous studies listed above, have been able to show that bars can
build small and close to exponential bulges (see also Courteau, de
Jong \& Broeils 1996 for observational evidence for bulge formation
out of disks). However, no simulations have been able to use bar
instabilities to produce very large bulges or even ellipticals.  On
the other hand, all these simulations started from a normal,
marginally unstable disk, whereas the process we are describing starts
from a highly unstable, compact disk, includes the continuous supply
of baryonic matter from prolonged cooling flows, and is likely to have
additional heating sources due to the star formation and feedback
processes which are expected to take place in this phase of the bulge
formation. Detailed numerical simulations that take all these
processes into account are required in order to investigate whether
this can indeed lead to massive bulges as assumed here.  Finally, we
emphasize that the high densities of the inner shells yield cooling
time-scales that may be significantly shorter than the dynamical
time-scale. This can add to the inside-out formation of a bulge
component, as suggested by Kepner (1997).

We assume that the bulge formation process discussed above
continuous until the bulge has become massive enough such that the
subsequent layers of baryonic material can cool and form a disk which
is stable against bar-formation. The process of disk-bulge formation
is thus a self-regulating one in that the bulge grows until it is
massive enough to sustain the remaining gas in the form of a stable disk.
Since the stability is directly related to the amount of angular
momentum of the gas (see Section~2.4), we expect a clear correlation
between the disk-to-bulge ratio and the angular momentum of the dark
halo out of which the galaxy is assembled.  We will show that this
correlation is present, and that it nicely follows the marginal
stability curve as expected from the self-regulating mechanism
proposed here.

\subsection{A disk-bulge-halo model} 
\label{s22}

The galaxies under investigation in this paper, varying from spirals
to disky ellipticals, are assumed to consist of a disk plus bulge
embedded in a dark halo. Virialized dark halos can be quantified by a
mass $M$, a radius $r_{200}$, and an angular momentum $J$ (originating
from cosmological torques).  Spherical collapse models (e.g., Gunn \&
Gott 1972) show that virialized halos have an over-density of
approximately 200.  In a recent series of papers Navarro, Frenk \&
White (1995; 1996; 1997, hereafter NFW) showed that virialized dark
halos have universal equilibrium density profiles, independent of mass
or cosmology (but see Moore \etal 1997), which can be well fit by
\begin{equation}
\label{nfwprof}
\rho(r) = \rho_{\rm crit} {\delta_0 \over (r/r_s) (1 + r/r_s)^2},
\end{equation}
where
\begin{equation}
\label{overdens}
\delta_0 = {200 \over 3} {c^3 \over {\rm ln}(1+c) - c/(1+c)},
\end{equation}
with $c=r_{200}/r_s$ the concentration parameter, $r_s$ a scale
radius, and $\rho_{\rm crit}$ the critical density for closure.  The
radius $r_{200}$ is defined as the limiting radius of a virialized
halo, and corresponds to the radius within which the mean density is
$200 \rho_{\rm crit}$.  Throughout we only consider spherical halos.
Rather than specifying the halo by its mass $M$, it is customary to
specify it by its circular velocity $V_{200} \equiv
\sqrt{GM/r_{200}}$.  Given the mass $M$, the formation redshift
$z_{\rm form}$, and the specific cosmology, the concentration
parameter $c$ can be calculated using the procedure outlined in
Appendix A of Navarro, Frenk \& White (1997). In this paper we compare
two different cosmologies: a standard cold dark matter cosmology
(SCDM) with a total matter density of $\Omega_0 = 1.0$, a Hubble
constant of $H_0 = 100 h \kms {\rm Mpc}^{-1}$ with $h=0.5$, and with a
rms linear over-density at $z=0$ in spheres of radius $8 h^{-1}$ Mpc of
$\sigma_8 = 0.6$. This latter parameter normalizes the CDM power
spectrum.  The second cosmology discussed here is an open cold dark
matter model (OCDM) with $\Omega_0 = 0.3$, $h=0.6$, and $\sigma_8 =
0.68$. Both models have a zero cosmological constant and the
parameters are typical of those favored by large-scale structure
constraints. The baryon density is set to be $\Omega_{\rm bar} = 0.0125 \,
h^{-2}$, in agreement with the nucleosynthesis constraints (e.g.,
Walker \etal 1991). The baryonic mass fraction is then given by
$\Omega_{\rm bar}/\Omega_0$.

We assume the final system to have a disk-to-bulge ratio $t \equiv
M_d/M_b$, such that the masses of disk, $M_d$, and bulge, $M_b$, can
be written as
\begin{equation}
\label{dismass}
M_d = \epsilon_{gf} \> {t \over 1+t} \> {\Omega_{\rm bar} \over \Omega_0} \>
M \equiv m_d M,
\end{equation}
and
\begin{equation}
\label{bulmass}
M_b = \epsilon_{gf} \> {1 \over 1+t} \> {\Omega_{\rm bar} \over \Omega_0} \>
M \equiv m_b M.
\end{equation}
Here $\epsilon_{gf}$ is the galaxy formation efficiency, expressed as
the fraction of the available baryons that ultimately end up in the
galaxy (see Section~2.3).  As mentioned in Section~2.1, we assume that
the bulge is formed out of the low-angular momentum material.
Furthermore, we assume that the material that builds the disk does
not loose its specific angular momentum.  Under these conditions we
have that $J_d \equiv j_d J$, where
\begin{equation}
\label{angmomdisk}
j_d = f(t,\epsilon_{gf}) [m_d + m_b].
\end{equation}
The function $f(t,\epsilon_{gf})$ depends on the disk-to-bulge ratio
and on the detailed physics that describe how the fraction
$\epsilon_{gf}$ of the baryonic material ends up in the disk-bulge
system (see Section~2.3). This function is derived in the Appendix. In
the following we parameterize $J$ by the dimensionless spin parameter
$\lambda$, defined by
\begin{equation}
\label{spinparam}
\lambda = {J \vert E \vert^{1/2} \over G M^{5/2}}.
\end{equation}
Here $E$ is the halo's total energy. Several studies, both analytical
and numerical, have shown that the distribution of spin angular
momentum of collapsed dark matter halos is well approximated by a log
normal distribution:
\begin{equation}
\label{spindistr}
p(\lambda){\rm d} \lambda = {1 \over \sigma_{\lambda} \sqrt{2 \pi}}
\exp\biggl(- {{\rm ln}^2(\lambda/\bar{\lambda}) \over 2
  \sigma^2_{\lambda}}\biggr) {{\rm d} \lambda \over
  \lambda},
\end{equation}
(e.g., Barnes \& Efstathiou 1987; Ryden 1988; Cole \& Lacey 1996;
Warren \etal 1992). We consider a distribution that peaks around
$\bar{\lambda} = 0.05$ with $\sigma_{\lambda} = 0.7$. These values are
in good agreement with the $N$-body results of Warren \etal (1992).

We assume the disk to be an infinitesimally thin exponential disk,
with surface brightness
\begin{equation}
\label{sigdisk}
\Sigma_d(R) = \Sigma_{d,0} \exp\bigl(-R/R_d).
\end{equation}
and total luminosity
\begin{equation}
\label{ldisk}
L_d = 2 \pi \Sigma_{d,0} R_d^2.
\end{equation}
In order to convert the disk parameters to an observers frame we need
to assume a mass-to-light ratio $\Upsilon_d$.  Bottema (1997) found
$\Upsilon_B = (1.79 \pm 0.48)$ for $h=0.75$ for HSB disks. Using $B-V
\approx 0.8$ (e.g., de Jong 1996b) and $(B-V)_{\odot} = 0.65$ this
yields $\Upsilon_d = 2.1 h$ in the $V$-band.

Although bulges and elliptical galaxies are reasonably well fit by
a $r^{1/4}$ `de Vaucouleurs' law, it has recently been shown that a
better fit is obtained by using a generalized $r^{1/n}$ law. This
fitting function, introduced by Sersic (1968), has the form
\begin{equation}
\label{sigbulge}
\Sigma_b(r) = \Sigma_{b,0} \exp\biggl[-a \Bigl({r \over 
r_e}\Bigr)^{1/n}\biggr],
\end{equation}
where $\Sigma_{b,0}$ is the central surface brightness, $r_e$ is the
effective radius, and $n$ is the exponential variable. For $n=1$ one
has a pure exponential, while for $n=4$ one recovers the de
Vaucouleurs profile.  The parameter $a$ is determined by the
requirement that $r_e$ is the radius encircling half of the total
luminosity and is very well approximated by
\begin{equation}
\label{abulge}
a = 2.0 n - 0.324 ,
\end{equation}
(Ciotti 1991).  The total luminosity of a system with a $r^{1/n}$
luminosity profile (assuming sphericity) is given by
\begin{equation}
\label{lbulge}
L_b = {n \Gamma (2n) \over a^{2n}} 2 \pi \Sigma_{b,0} r_e^2,
\end{equation}
where $\Gamma$ is the Gamma function. It was found that the best
fitting exponent $n$ is correlated with the total luminosity of the
spheroid: Andredakis, Peletier \& Balcells (1995) used the
$r^{1/n}$ law to fit the luminosity profiles of bulges of spiral
galaxies, and found the effective radius $r_e$ to be related to the
total luminosity of the bulge by the empirical relation
\begin{equation}
\label{mbulge}
M_B = -19.75 - 2.8 \, {\rm log}(r_e),
\end{equation}
In addition, Caon, Capaccioli \& D'Onofrio (1993) found that the
exponent $n$ is correlated with the effective radius through the
empirical relation
\begin{equation}
\label{nbulge}
{\rm log}(n) = 0.28 + 0.52 \, {\rm log}(r_e).
\end{equation}
Although this empirical relation has only been shown to fit early-type
galaxies, we assume here that it is also valid for bulges.  We take
$n$ to be limited between $n=1$ (exponential) and $n=4$ (de
Vaucouleurs). None of our results are significantly influenced by this
assumption however. The main parameter for the bulge is its total
mass; changes in its actual density distribution are only a second order
effect.  The relations (\ref{sigbulge}) -- (\ref{nbulge}) allow us
to describe the bulges by a single parameter; once the total
luminosity of the bulge is known, we can use the empirical relations
to determine all other bulge parameters!  Van der Marel (1991) found
that ellipticals have $\Upsilon_R = 6.64 h$ (Johnson $R$).  Using $V -
R_J \approx 0.68$ (e.g., van der Marel 1991) and $(V-R_J)_{\odot} =
0.52$ this yields $\Upsilon_b = 7.7 h$ in the $V$-band. For our
combined disk-bulge model we assume that the mass-to-light ratio is
constant over the entire galaxy with a value equal to the luminosity
weighted average of the mass-to-light ratios of the disk and bulge.
Alternatively, we could have assigned the bulge and disk separate
mass-to-light ratios of $\Upsilon_d$ and $\Upsilon_b$ respectively,
but the mass-to-light ratio of the disks in S0s and disky ellipticals
is clearly larger than for spiral discs. Assuming a constant,
luminosity weighted ratio is thus more realistic.

Once the baryons inside the virialized dark halo start to cool, this
induces a contraction of the inner region of the dark halo.  We follow
Blumenthal \etal (1986) and Flores \etal (1993) and assume that the
halo responds adiabatically to the slow assembly of the baryonic
matter in disk and bulge, and that it remains spherical (see also
Dalcanton \etal 1997 and Mo \etal 1998). Under these assumptions the
angular action of the dark halo material is an adiabatic invariant,
such that a particle initially at a radius $r_i$ ends up at mean
radius $r$ according to
\begin{equation}
\label{initialrad}
r_i \, M_i(r_i) = r \, M_f(r).
\end{equation}
Here $M_i(r)$ is the initial mass distribution (the NFW profile)
given by 
\begin{equation}
\label{initmassfunc}
M_i(r) = M \Biggl[{{\rm ln}(1 + cx) - cx/(1+cx) \over
                     {\rm ln}(1 + c) - c/(1+c)}\Biggr],
\end{equation}
with $x=r/r_{200}$, and $M_f(r)$ is the final mass distribution given by
\begin{equation}
\label{finalmassfunc}
M_f(r) = M_d(r) + M_b(r) + (1-m_d-m_b)M_i(r_i).
\end{equation}
Here
\begin{equation}
\label{diskmassfunc}
M_d(r) = M_d \biggl[1 - \Bigl(1 + {r \over R_d}\Bigr) \exp(-r/R_d)\biggr],
\end{equation}
and
\begin{equation}
\label{bulmassfunc}
M_b(r) = {2 a^{2n+1} \over \pi n^2 \Gamma (2n)} M_b
\int\limits_{0}^{r/r_e} {\rm d}y \, y^2 \int\limits_y^{\infty}
x^{\beta - 1} \exp\bigl[-a x^{\beta}\bigr] {{\rm d}x \over \sqrt{x^2 -
    y^2}},
\end{equation}
with $\beta = 1/n$.

Mo \etal (1998) have shown that the scale-length of the disk can be 
written as
\begin{equation}
\label{rdlambda}
R_d = {1 \over \sqrt{2}} \Bigl({j_d\over m_d}\Bigr)
\lambda \, r_{200} f_c^{-1/2} f_R^{-1},
\end{equation}
where
\begin{equation}
\label{frint}
f_R = {1\over 2} \int_0^{\infty} u^2 e^{-u} {V_c(R_d u) \over V_{200}}
{\rm d}u,
\end{equation}
a parameter that indicates the amount of self-gravity of the disk, and
\begin{equation}
\label{fcnfw}
f_c = {c \over 2} {1 - 1/(1+c)^2 - 2 \, {\rm ln}(1+c)/(1+c) \over
[c/(1+c) - {\rm ln}(1+c)]^2}.
\end{equation}
The circular velocity $V_c(r)$ is given by
\begin{equation}
\label{vctot}
V_c^2(r) = V_{c,d}^2(r) + V_{c,b}^2(r) + V_{c,DM}^2(r),
\end{equation}
with
\begin{equation}
\label{vrotbul}
V_{c,b}^2(r) = {G M_b(r) \over r},
\end{equation}
$V_{c,d}^2(r)$ as given by equation [2.196] in Binney \& Tremaine
(1987), and
\begin{equation}
\label{vrothalo}
V_{c,DM}^2(r) = {G \bigl[M_f(r) - M_d(r) - M_b(r)\bigr] \over r}.
\end{equation}

We can solve this set of equations in an iterative way, following the
procedure used by Mo \etal (1998). For a given $M$, $\lambda$, $z_{\rm
  form}$ and $t$ we first calculate the concentration parameter $c$ of
the dark halo before collapse of the baryons. We then calculate the
parameters of the bulge using equations (\ref{abulge}) --
(\ref{nbulge}).  Next we start with a guess for $f_R$, by setting $f_R
= 1$, and we calculate the disk parameters $R_d$ and $\Sigma_{0,d}$
using equations (\ref{rdlambda}) and (\ref{ldisk}) respectively.
Subsequently we solve for $r_i$ as a function of $r$ using equations
(\ref{initialrad}) -- (\ref{bulmassfunc}), and thus obtain $M_f(r)$.
We then calculate $V_c^2(r)$, and substitute this in equation
(\ref{frint}) to obtain a new value for $f_R$.  Convergence is rapidly
achieved: 3 to 5 iterations yield an accuracy of $R_d$ already better
than one percent.

\subsection{Estimating the galaxy formation efficiency}

Since we have defined the mass of the dark halo such that the average
over-density of the halo is 200 times the critical density for closure
at the virialization redshift $z$, we have that
\begin{equation}
\label{massvir}
M = {V_{200}^3 \over 10 G H(z)}.
\end{equation}
This halo mass can be related to the disk luminosity (assuming no bulge,
i.e., $t=\infty$), by
\begin{equation}
\label{masseff}
M = {\Upsilon_d L_d \over \epsilon_{gf} (\Omega_{\rm bar}/\Omega_0)},
\end{equation}
(cf. equation [\ref{dismass}]). Combining the above equations yields
\begin{equation}
\label{modtf}
L_d = {\epsilon_{gf} (\Omega_{\rm bar}/\Omega_0) \over 10 G \Upsilon_d H(z)}
V_{200}^3. 
\end{equation}
This is similar to the Tully-Fisher relation: $L = A
(V_c/220)^{\gamma}$, whereby the zero-point, $A$, is inversely
proportional to $H(z)$. However, current data suggest no zero-point
evolution (Vogt \etal 1996, 1997; Mao, Mo \& White 1997).  In other
words, if we assume a significant scatter in the formation redshifts
of spiral galaxies (some massive spirals are already in place at $z
\sim 1$, see Vogt \etal 1996), the dependence of equation
(\ref{modtf}) would yield a scatter in the Tully-Fisher relation much
larger than observed: if spirals form between $z=0$ and $z=1$ (and
they are not significantly modified since their formation) a scatter
of $\sim 1.1$ mag is predicted for an Einstein-de Sitter Universe,
much larger than the observed scatter of $\lesssim 0.3$ mag. This suggests
that the redshift dependence of the Tully-Fisher zero-point has to be
corrected for by either imposing $\Upsilon_d \propto H(z)^{-1}$ or
$\epsilon_{gf} \propto H(z)$.  We consider the former option unlikely,
since this predicts spirals at $z \sim 1.0$ to be much bluer than
observed (see Vogt \etal 1996). We therefore consider a scenario in
which the galaxy formation efficiency is redshift dependent. Combining
the observed Tully-Fisher relation with equation (\ref{modtf}) yields
\begin{equation}
\label{diskeff}
\epsilon_{gf} = 0.678 \> h^2 \> \Omega_0 \> \biggl({A\over
  L^{*}}\biggr)^{3/\gamma} \biggl({L\over L^{*}}\biggr)^{(1-3/\gamma)}
{H(z) \over H_0},
\end{equation}
where $L^{*} = 1.0 \times 10^{10} \, h^{-2} \Lsun$.  Using the
Tully-Fisher relation of Giovanelli \etal (1997), converted to the
$V$-band using $V-I = 1.0$ as an average for spirals (de Jong 1996b),
yields
\begin{equation}
\label{diskefb}
\epsilon_{gf} = 1.562 \> h^2 \> \Omega_0 \> {H(z)\over H_0},
\end{equation}
(i.e., Giovanelli \etal find $\gamma \approx 3.0$). Here we 
assumed that the HI-linewidth equals twice the circular velocity
$V_{200}$.

The important factor here is 
\begin{equation}
\label{hubconst}
{H(z)\over H_0} = \sqrt{\Omega_{\Lambda} + (1 - \Omega_{\Lambda} -
  \Omega_0) (1+z)^2 + \Omega_0 (1+z)^3},
\end{equation}
which shows that the galaxy formation efficiency depends on the
specific cosmology and the formation redshift. Here $\Omega_{\Lambda}
= \Lambda / (3 H_0^2)$, with $\Lambda$ the cosmological constant.  In
the SCDM model with $h=0.5$ and $\Omega_0 = 1.0$, equation
(\ref{diskefb}) yields $\epsilon_{gf} = 0.39 (1+z)^{3/2}$.  Galaxies
that were assembled at higher redshift thus formed more efficiently.
More efficiently here means that a larger fraction of the baryonic
matter available in the dark halo actually made it into the disk-bulge
system. At $z=0$ this fraction is approximately forty percent, and
this increases to hundred percent at $z \geq 0.87$. Halos of a given
mass are denser at higher redshift, and higher density implies higher
escape velocity and higher cooling efficiency, both of which could
give a plausible explanation for an increase of the galaxy formation
efficiency with increasing redshift.

In what follows we consider two extreme cases of this galaxy formation
(in)efficiency. In the first scenario, which we call the
`cooling'-scenario, we assume that only the inner fraction
$\epsilon_{gf}$ of the baryonic mass is able to cool and form the
disk-bulge system: the outer parts of the halo, where the density is
lowest, but which contain the largest fraction of the total angular
momentum, never gets to cool. In the second scenario, referred to
hereafter as the `feedback'-scenario, we assume that the feedback from
star formation and evolution is such that each baryon in the dark
halo, independent of its specific angular momentum, has a probability
of $\epsilon_{gf}$ to ultimately end up as a constituent of the
disk or bulge.  The choice of either of these scenarios enters our
equations in the function $f(t,\epsilon_{gf})$ (see the Appendix).

\subsection{Stability}

As is well known, disks that are self-gravitating tend to be unstable
against bar formation. Halos with smaller $\lambda$ yield disks with
smaller scale-lengths (see equation [\ref{rdlambda}]). For a given
disk-to-bulge ratio, there is thus a critical $\lambda$ such that for
$\lambda < \lambda_{\rm crit}$ the disk will be bar unstable.
Christodoulou, Shlosman \& Tohline (1995) derived a stability
criterion for bar formation based on the angular momentum content of
the disk, rather than the energy content (such as the well known
Ostriker \& Peebles (1973) criterion).  For a disk-bulge-halo system
this stability criterion can be written such that disks are stable
against bar formation if
\begin{equation}
\label{stabalpha}
\alpha = {V_d \over 2 V_c} < \alpha_{\rm crit},
\end{equation}
with $V_d$ and $V_c$ the characteristic circular velocities of the
disk and the composite disk-bulge-halo system (equation [\ref{vctot}]),
respectively.  This criterion is similar to the one proposed by
Efstathiou, Lake \& Negroponte (1982). As characteristic velocities we
consider the circular velocities at $R = 3 R_d$, since Mo \etal (1998)
have shown that this is the radius where typically the rotation curve
of the composite system reaches a maximum. For a thin exponential
disk, approximately 80 percent of the disk mass is located inside $3
R_d$, and
\begin{equation}
\label{vdapprox}
V_d^2(3 R_d) \approx 0.359 {G M_d \over R_d}.
\end{equation}
Substituting equation (\ref{vdapprox}) in (\ref{stabalpha}), and using
equation (\ref{rdlambda}) we find that disks are stable if the spin
parameter of their halos obeys
\begin{equation}
\label{stablambda}
\lambda > \sqrt{2} \varepsilon_{\rm crit}^2 {m_d^2\over j_d} 
\biggl({V_{200} \over V_c(3R_d)}\biggr)^2 f_c^{1/2} f_R 
\equiv \lambda_{\rm crit},
\end{equation}
with
\begin{equation}
\label{epsstab}
\varepsilon_{\rm crit} = {\sqrt{0.359} \over 2 \alpha_{\rm crit}}.
\end{equation}
This reduces to $\varepsilon_{\rm crit} = 1.15$ for stellar disks and
$\varepsilon_{\rm crit} = 0.86$ for gaseous disks, using $\alpha_{\rm
  crit} = 0.26$ and $\alpha_{\rm crit} = 0.35$, respectively (see
Christodoulou \etal 1995). This is in excellent agreement with the
empirical value of $\varepsilon_{\rm crit} = 1.1$ found by Efstathiou
\etal (1982) based on $N$-body experiments.  Throughout we take
$\varepsilon_{\rm crit} = 1.0$ to take into account that disks are often
made up of a mix of gas and stars.

The critical value of $\lambda$ given by equation (\ref{stablambda})
depends on the galaxy formation efficiency $\epsilon_{gf}$, and thus
on the specific cosmology and formation redshift according to equation
(\ref{hubconst}). Furthermore, since $\lambda_{\rm crit} \propto
j_d^{-1}$, and $j_d$ is different for the cooling and feedback
scenarios, $\lambda_{\rm crit}$ depends on the particular model we
choose for the galaxy formation (in)efficiency.  In Figure~1 this
dependence is shown for three different cosmologies: SCDM, OCDM, and
$\Lambda$CDM. The latter is a model with a non-zero cosmological
constant with $\Omega_0 = 0.3$, $\Omega_{\Lambda} = 0.7$, $h=0.6$ and
$\sigma_8 = 0.97$.  Results are shown for both the feedback scenario
(left panel) and the cooling scenario (right panel).  Figure~1 plots
the logarithm of the critical spin parameter versus redshift.  The
dotted lines outline the probability distribution of the spin
parameter (equation [\ref{spindistr}]); the indices (indicated in the
right panel only) denote the probability of finding a halo below that
line. At low formation redshifts $\lambda_{\rm crit}$ increases with
redshift, due to the increasing galaxy formation efficiency with
redshift.  The turnovers to a flat line at higher redshift are caused
by the requirement that $\epsilon_{gf} \leq 1.0$. Note that in a SCDM
cosmology with $z_{\rm form} \geq 0.87$ (for which $\epsilon_{gf} =
1.0$) about 50 percent of all halos yield unstable disks, and thus
require the formation of a bulge component in order to stabilize the
disk.  This increases to $\sim 90$ percent in models with $\Omega_0 =
0.3$.  Furthermore, it is apparent that $\lambda_{\rm crit}$ is much
less redshift dependent in the case of the cooling scenario, than in
the case of the feedback scenario. The implications of this are
discussed in Section~4.  As can be seen, the results are very similar
for the two low $\Omega_0$ models. Since $\lambda_{\rm crit}$ is the
important parameter for what follows, we only focus on a comparison of
SCDM with OCDM cosmologies, and note that results for the $\Lambda$CDM
model are similar as for the OCDM model.

\subsection{Isothermal halo profiles}

It has recently been argued that the results of Navarro, Frenk \&
White concerning the universal density profiles of dark halos may be
incorrect.  Very high resolution simulations by Moore \etal (1997)
indicate that the central density cusp of dark halos may be even
steeper than $r^{-1}$. We therefore investigate the sensitivity of
our results to the assumption of a NFW halo profile by comparing our
results with those obtained considering isothermal halo profiles,
i.e., halos with a density distribution given by
\begin{equation}
\label{rhoiso}
\rho(r) = {V_c^2 \over 4 \pi G r^2}.
\end{equation}
For such halo profiles the circular velocity, $V_c$, is independent of
radius and equal to the circular velocity at the virial radius,
$V_{200}$.  The results for the isothermal density profiles can be
obtained from the equations in this paper by taking $f_c = 1$.  We
have rerun all models with an isothermal halo, and compared the
results with those of the NFW halo profile. None of the results
presented in this paper are significantly influenced by the assumption
of a NFW halo profile.

\section{Towards understanding the $\mu_0$--$R_d$ diagram}

One of the strongest suggestions for a smooth transition from spirals
and S0s towards disky ellipticals along a sequence of decreasing
disk-to-bulge ratio is provided by the $\mu_0$--$R_d$ diagram, in
which the central surface brightness of disks, $\mu_0$, is plotted
against the disk's scale-length $R_d$ (see Scorza \& Bender 1995).
Scorza \& van den Bosch (1998) investigated the continuity of disk
properties of a variety of disk galaxies and extended the diagram
towards much higher $\mu_0$ and smaller $R_d$ by including the nuclear
disks discovered from Hubble Space Telescope photometry in a number of
early-type galaxies (van den Bosch \etal 1994; Kormendy \etal 1996;
van den Bosch, Jaffe \& van der Marel 1997). This extended
$\mu_0$--$R_d$ diagram, already published in Scorza \& van den Bosch
(1998), is shown in Figure~2. It plots the central surface brightness
(in $V$-band) as a function of the logarithm of the scale-length of a
vast variety of disks, ranging from the extraordinary large disk of
Malin~I (data taken from Bothun \etal 1987), to the very small nuclear
disks (data taken from Scorza \& van den Bosch 1998).  In addition,
Figure~2 plots the parameters of embedded disks (solid circles, data
taken from Scorza \& Bender 1995 and Scorza \etal 1998), and of a
combined sample of S0s (stars), HSB disks (open circles), and LSB
disks (triangles).  Data are taken from Kent (1985), de Jong (1996a),
Sprayberry \etal (1995), de Blok, van der Hulst \& Bothun (1995), and
McGaugh \& Bothun (1994).  Figure~2 combines the parameters of disks
that span 4 orders of magnitude in both scale-length and central
surface brightness!
  
Although there is considerable `scatter', the $\mu_0(V)$--$R_d$
diagram of Figure~2 seems to reveal an almost linear relation. It is
tempting to believe that this may teach us something about the
formation of galactic disks. However, it should be noted that
selection effects play an important role at the lower left part of the
diagram: Most of the spirals and S0s in Figure~2 are taken from large
diameter limited surveys, and are thus strongly biased towards the
largest scale-lengths. Furthermore, one is always strongly biased
against low surface brightness disks, especially if they are embedded
in a much brighter spheroid.  Even more, the fact that selection
effects are different for the various data sets combined in Figure~2
prohibits a statistical conclusion on this lower left part of the
diagram (e.g., see detailed discussions in de Jong 1996b and Dalcanton
\etal 1997).  However, the absence of disks in the upper right part of
the diagram is real, and should in principle be related to the
formation and evolution of disk systems.

We now use our disk-bulge-halo models to investigate the locations of
the different sorts of disks in the $\mu_0(V)$--$R_d$ diagram.  First
we note that spirals and S0s with the highest central surface
brightness for given scale-length lie approximately along a line of
constant disk luminosity. Part of the upper boundary in the
$\mu_0(V)$--$R_d$ diagram thus seems related to a cut-off in disk
luminosity. This is to be expected from the cut-off in the galaxy
luminosity function, which is well described by the Schechter (1976)
function. This tells us that there are only very few galaxies with
luminosities exceeding $3 L^{*}$. We thus assume that the upper
boundary is a line along which the sum of bulge and disk luminosities
is equal to $3 L^{*}$.  The dotted line in Figure~2 (which is partly
drawn as a solid thick line) corresponds to $L_d = 3 L^{*}$. The
position of a disk along a line of constant $L_d$ is determined by the
spin parameter $\lambda$: disks originating from halos with smaller
$\lambda$ have smaller scale-lengths, and higher central surface
brightness. As shown in Section~2.4, there is a critical spin
parameter such that for $\lambda < \lambda_{\rm crit}$ disks are
unstable. The disk for which $L_d = 3 L^{*}$ and $\lambda =
\lambda_{\rm crit}$ is indicated by a thick solid dot in Figure~2.  If
we want to build disks with scale-lengths smaller than this, we need a
bulge to help stabilize the disk.  The curved thick solid line, that
originates at the thick solid dot and extends towards smaller
scale-lengths, corresponds to a line with $L_d + L_b = 3 L^{*}$,
$\lambda = \lambda_{\rm crit}$, and with a disk-to-bulge ratio $t$
running from $t=\infty$ (big solid dot) towards $t=0$. The dashed line
is the line with $\lambda = \lambda_{\rm crit}$, $t=\infty$, and $L_d$
running from $3 L^{*}$ (big solid dot) to $L_d = 0$.  Using these
limits on the expected scale-lengths and central surface brightnesses
of disks in our disk-bulge-halo models, we can distinguish 4 different
regions in the $\mu_0(V)$--$R_d$ diagram, labeled I to IV in Figure~2.
Region I is the region where the disk luminosity exceeds $3 L^{*}$.
Region II is the region where $L_d < 3 L^{*}$ and $\lambda >
\lambda_{\rm crit}$; disks in this region are stable against bar
formation even without a bulge.  Disks in region III, however, require
a bulge to be stable. Region IV, finally, is the region where systems
with $L_d + L_b < 3 L^{*}$ have an unstable disk despite the presence
of the bulge.  The position of the big solid dot, and therewith of the
boundary between regions II and III, depends strongly on the formation
redshift and cosmology (cf.  Figure~1). Figure~2 is based on a SCDM
cosmology with zero formation redshift. Although quite a number of
assumptions have gone into our disk-bulge-halo models, the upper
boundary (indicated by the thick solid line) nicely separates the
region without disks, from the region most densely occupied by disks.
Of course, the fact that the number density is highest just below the
boundary is an observational bias effect as discussed above.  One has
to keep in mind that the $\mu_0$--$R_d$ diagram can be rather
confusing since there are {\it three} parameters that determine the
position of a disk in this diagram, namely the disk luminosity, the
spin parameter, and the disk-to-bulge ratio. In the following we
therefore focus on another diagram in which the luminosity dependence
has been removed.

\section{Clues to the formation of bulge-disk systems}

We now ask ourselves whether we can learn something about the
formation of disks and bulges from a comparison of our disk-bulge-halo
models with real galaxies.  We do this by calculating, for each galaxy
in our combined sample, the spin parameter $\lambda$ of the dark halo
which, under a given set of assumptions, yields the observed disk
properties (scale-length and central surface brightness).  We thus use
our formation scenario to link the {\it disk} parameters to those of
the dark halo, and use the statistical properties of dark halos to
discriminate between different cosmogonies. We start by choosing a
cosmology and a formation redshift and we read in the disk-to-bulge
ratio, $t$, and the scale-length and central surface brightness of the
disk. From this we determine the luminosities of the disk and bulge,
which, upon using equations (\ref{dismass}), (\ref{bulmass}), and
(\ref{diskefb}), yield the total mass $M$ and the particular density
distribution of the disk-bulge-halo system.  We then calculate
$\lambda$ from equation (\ref{rdlambda}) and plot this against the
disk-to-bulge ratio $t$.  This $\lambda$ is the value of the spin
parameter that the halo would have had if the observed disk-bulge
system formed out of that halo in the way envisioned here. One of the
advantages of this diagram is that the dependence on luminosity is
eliminated: the position of a disk in this diagram is virtually
independent of the disk's total luminosity.

In Figures~3 (SCDM) and~4 (OCDM) we plot the location of disks of LSB
spirals (triangles), HSB spirals (open circles), S0s (stars), and
disky ellipticals (solid circles) in a $\log[\lambda]$ vs. $\log[\, t
\,]$ diagram.  The five small-dashed lines in Figures~3 and~4
correspond to the value of $\lambda$ below which we expect $1$, $10$,
$50$, $90$, and $99$ percent of the halos for a distribution function
of spin parameters given by equation (\ref{spindistr}) with
$\bar{\lambda} = 0.05$ and $\sigma_{\lambda} = 0.7$ (see also
Figure~1). These lines are drawn to guide the eye.  If our assumptions
for the formation of disk-bulge-halo systems are correct, the
distribution of inferred halo spin parameters should be similar to
this log normal distribution.

The thick solid lines in Figures~3 and~4 correspond to $\lambda =
\lambda_{\rm crit}$ (equation [\ref{stablambda}]). Disks that lie
below this line are unstable against bar formation.  This critical
value of the spin parameter has been calculated for a halo with a
total mass of $10^{13} \Msun$, but is virtually identical for other
halo masses.  The crowding of disks at $t=10$ is due to the fact that
we have assigned a disk-to-bulge ratio of ten to systems for which no
value of $t$ was determined (these are mainly a number of the LSB
spirals).  The panels on the left correspond to $z_{\rm form} = 0.0$,
the panels in the middle to $z_{\rm form} = 1.0$, and the panels on
the right to $z_{\rm form} = 3.0$. The upper three panels correspond
to models in which $\epsilon_{gf} = 1.0$, i.e., in which {\it all} the
baryons in the dark halo make it into the galaxy.  The panels in the
middle correspond to the cooling scenario, and the lower panels to the
feedback scenario. The main difference between the upper and the
middle and lower panels is that the former have $\lambda_{\rm
  crit}$ independent of formation redshift (there is a non-zero but
negligible dependence on $z_{\rm form}$ left due to the fact that the
concentration parameter $c$ of the NFW halo profiles depends on
formation redshift), whereas $\lambda_{\rm crit}$ increases with
$z_{\rm form}$ in the middle and lower panels (according to the curves
in Figure~1).

Since $R_d \propto r_{200}$ (equation [\ref{rdlambda}]) and $r_{200} =
0.1 V_{200}/H(z)$, disks forming in a halo with mass $M$ and spin
parameter $\lambda$ are smaller for higher $z_{\rm form}$. A higher
formation redshift thus implies a larger spin parameter in order to
yield the observed value of $R_d$, something which is evident from
Figures~3 and~4. Furthermore, disks seem to follow the $\lambda =
\lambda_{\rm crit}$ curve, i.e., disk-bulge systems do not have bulges
that are significantly more massive than required by disk-stability.
This is at least consistent with the inside-out bulge formation scenario
proposed in Section~2.1, in which the formation of disk and bulge is
self-regulating and governed by disk stability.

The OCDM model with constant and near to unity galaxy formation
efficiency (upper panels in Figure~4) can be ruled out. In this
scenario systems with a large disk-to-bulge ratio, such as spirals,
are rare; only $\sim 10$ percent of the dark halos is expected to
yield such systems. The stability line $\lambda = \lambda_{\rm crit}$
crosses the average $\bar{\lambda}$ of the spin parameter distribution
at $D/B \approx 1.0$, and thus half of the halos is expected to form
systems with a disk-to-bulge ratio less than unity.  This is in
conflict with the fact that the major fraction of field galaxies are
spirals. Furthermore, stable spirals can only form at high redshift
($z_{\rm form} \gta 3$), since for lower formation redshifts the
spirals fall below the stability line. In the OCDM cooling scenario,
spirals that formed recently tend to be slightly more stable, but
still, the probability that a certain halo yields a system with a
large disk-to-bulge ratio is rather small, rendering this scenario
improbable. However, for the feedback OCDM cosmogony, a promising
scenario unfolds: proto-galaxies that collapse at high redshifts are
preferentially found in overdense regions such as clusters. The
densities of these virialized halos are high, and this (somehow)
causes the galaxy formation to be very efficient; nearly all the
baryons in the dark halos make it into the galaxy. As a consequence,
the systems that form require relatively small disk-to-bulge ratios
(e.g., similar to S0s) in order to be stable.  As the Universe
evolves, and proto-galaxies with smaller over-densities (which are
predominantly located in the field) start to turn around and collapse,
the galaxy formation efficiency decreases, and systems with larger
disk-to-bulge ratios are allowed to form.  This scenario thus
automatically yields a morphology-density relation, in which systems
with smaller disk-to-bulge ratios (e.g., S0s) are preferentially to be
found in overdense regions.  Unfortunately, the disky ellipticals seem
to not fit in this scenario.  Even at formation redshifts of three,
the average disky elliptical lies near the line with $p(< \lambda) =
0.01$.

The scenario outlined above for our OCDM cosmology also applies to the
SCDM model with $\Omega_0 = 1$, albeit somewhat less restricting.
After all, $\lambda_{\rm crit}$ only gets as large as $\sim
\bar{\lambda}$. Consequently, in a SCDM Universe our model predicts
less of a morphology-density relation than in a low density Universe;
i.e., the magnitude of the morphology-density relation depends on the
baryon fraction $\Omega_{\rm bar}/\Omega_0$.

\subsection{Formation redshifts}
 
We define the typical formation redshift of a subset of galaxies as
that redshift for which the average disk yields $\lambda =
\bar{\lambda}$.  In Figure~5 we plot histograms of inferred formation
redshifts for spirals (HSB and LSB disks combined), S0s, and disky
ellipticals. The three dotted lines are drawn to guide the eye and
correspond to redshifts of 1, 3, and 10.  The inferred formation
redshift is determined by calculating the redshift at which a halo
with $\lambda = \bar{\lambda} = 0.05$ yields the observed disk
parameters. If a disk lies already above the $\lambda = 0.05$ line at
$z = 0$, a formation redshift of zero is assigned. The arrows in
Figure~5 indicate the mean of the inferred formation redshifts for
each sample.  As can be seen, spiral disks are consistent with having
formed fairly recently: $\langle z_{\rm form} \rangle \approx 0.3$
with a mean formation redshift of zero.  This conclusion was already
reached by Mo \etal (1998) based on the same disk formation scenario
as used here, but on somewhat different arguments. The distribution of
$z_{\rm form}$ is much more spread out for S0s, with a mean formation
redshift of $\sim 7$.  Note that misclassifications may be partially
responsible for this large spread.  Finally, the inferred formation
redshifts for the disky ellipticals are unrealistically high, with a
mean of $\sim 38$. We thus conclude that unless there is a significant
loss in specific angular momentum of the material that forms these
systems, disky ellipticals seem to {\it not} form a continuous
sequence with S0s and spirals, at least not in the sense of the
formation scenario investigated here.
  
The histograms in Figure~5 correspond to the feedback SCDM cosmogony.
However, the results are very similar for the other cosmogonies
discussed in this paper: although the details of the histograms are
somewhat different, the averages are the same within the errorbars.

\section{Alternative formation scenarios for disky ellipticals}

As discussed above it seems unlikely that disky ellipticals can be fit
in the simple bulge-disk formation scenario envisioned here. The small
disk-to-bulge ratios in these systems renders an inside-out formation
scenario for the bulges in disky ellipticals unlikely. Two alternative
formation scenarios for disky ellipticals have been proposed: (i)
disky ellipticals are the outcome of gas-rich mergers (e.g., Bender,
Burstein \& Faber 1992; Scorza \& Bender 1996), and (ii) embedded
disks are formed out of internal mass loss from the old stars in the
spheroid (Scorza 1993; Brighenti \& Mathews 1997).

Hernquist \& Barnes (1991) have shown how in gas-rich mergers the gas
becomes segregated in two components. The first component looses most
of its specific angular momentum and collapses to form a dense central
cloud.  When converted into stars, these compact blobs of gas may well
account for the high central densities in disky ellipticals (Barnes
1998). The second component gets propelled to larger radii in the
tidal tails originating from the merger. This gas may either escape or
reaccrete (through cooling) onto the newly formed `elliptical' and
form a new disk component, resembling the embedded disks. The numerous
physical processes at play and the large parameter space available
make it difficult to come up with detailed predictions for the
resulting merger remnant.  However, at present there seem to be no
clear observations that rule out gas-rich mergers as the origin for
disky ellipticals (but see Kissler-Patig, Forbes \& Minniti 1998).

In the internal mass loss scenario one expects a clear correlation
between the mass of the disk and the mass of the spheroid.  This
correlation is indeed present and Scorza \& Bender (1998) showed that
typical mass loss rates for the old spheroid population are in
excellent agreement with the typical ratio between the masses of the
embedded disk and the spheroid. Another prediction from this scenario
is that the spheroid and disk rotate in the same direction and that
their specific angular momenta are similar (at least if we assume that
the gas blown into the interstellar medium by the spheroid does not
loose its specific angular momentum when cooling to form the embedded
disk).  Scorza \& Bender (1995) found this to be the case from
kinematic decompositions of line-of-sight velocity profiles of a
number of disky ellipticals from which they determined estimates of
the specific angular momenta of the embedded disk and the spheroid.
Their data, which we plot in Figure~6, yields an average ratio of
\begin{equation}
\label{angmomrat}
\Bigl\langle {(J/M)_d \over (J/M)_b} \Bigr\rangle = 1.1 \pm 0.6.
\end{equation}
A third prediction associated with this scenario is that the disks are
younger than the spheroids. Recently, de Jong \& Davies (1997)
reported that disky ellipticals have higher H$\beta$ line indices than
the more luminous boxy ellipticals, and suggested this to be due to a
young population, most likely the embedded disk (see also Brighenti \&
Mathews 1997). A final prediction for the internal mass loss scenario
is that the disks have surface brightness profiles that are steeper
than exponential, since the gas does not cool from a uniform sphere in
solid body rotation, but from a differentially rotating, centrally
concentrated gas distribution (Scorza 1993; Brighenti \& Mathews
1997).  Indeed, Scorza \& Bender (1995) and Scorza \etal (1998) have
found that the embedded disks often have surface brightness profiles
that are significantly steeper than exponential.

The masses, specific angular momenta, surface brightness profiles, and
ages of embedded disks are thus all consistent with them having formed
out of mass loss from the old stellar population of the spheroid.
More accurate determinations of the specific angular momenta of
embedded disks and their spheroids in a larger sample of disky
ellipticals will proof helpful in constraining this further.

\section{Conclusions \& Discussion}

Understanding galaxy formation is intimately linked with understanding
the origin of the Hubble sequence. An important clue is provided by
comprehending the formation of disky ellipticals and S0s, simply
because these systems form the transitional class from the classical
ellipticals to the spirals. Disky ellipticals seem to form a
continuous sequence with spirals and S0s towards smaller disk-to-bulge
ratio, and it is thus tempting to believe that disky ellipticals, S0s
and spirals all formed in a similar fashion. The aim of this paper has
been to investigate a simple formation scenario for disk-bulge-halo
systems, and to search for the main parameters and/or processes that
determine the disk-to-bulge ratio and thus explain to a large extent
the origin of (the major part of) the Hubble sequence.

We considered the disk formation scenario originally proposed by Fall
\& Efstathiou (1980), in which the size of the disk, which is formed
by cooling of the gas in a dark halo, is determined by the amount of
angular momentum of the halo. We have extended upon this formation
scenario by including a simple picture for the inside-out formation
of an additional bulge component out of the inner, low angular
momentum material. Stability arguments are used to
suggest that the formation of the bulge is a self-regulating process
in which the bulge grows until it is massive enough to allow the
remaining gas to form a stable disk component.  We do not describe the
bulge formation in any detail but merely use empirical relations which
allow us to describe the bulge by a single parameter, namely its mass.

Each dark halo contains a fraction $\Omega_{\rm bar}/\Omega_0$ of baryons.  We
introduced a galaxy formation efficiency $\epsilon_{gf}$ which
describes the fraction of those baryons that actually build up the
disk-bulge system. The theory of spherical collapse coupled with the
definition of a virialized halo predicts a Tully-Fisher relation of
the form $L \propto V_c^3$ as observed, with a zero-point that depends
on the Hubble constant $H(z)$. Recent observations, however, suggest
that the Tully-Fisher zero-point is independent of redshift, implying
that the galaxy formation efficiency is proportional to $H(z)$.  A
physical explanation for this redshift dependence may be the higher
escape velocities and cooling efficiencies at higher redshifts.

For a combined sample of $\sim 200$ galaxies, varying from spirals to
disky ellipticals, we calculated the value of the halo's spin
parameter which yields the observed disk properties under the
assumption that disk-bulge systems form in the way envisioned here.
We compared two cosmologies (SCDM vs. OCDM) and investigated the
differences between assuming a galaxy formation efficiency of unity
and two (extreme) scenarios in which $\epsilon_{gf} \propto H(z)$: a cooling
scenario, in which we assume that only the inner fraction
$\epsilon_{gf}$ of the available baryons cools to form the disk and
bulge, and a feedback scenario, in which each baryon, independent of
its specific angular momentum, has a probability $\epsilon_{gf}$ of
ultimately ending up in the disk or bulge. Our main conclusions are
the following:

\begin{itemize}

\item Disk-bulge systems do not have bulges that are significantly
  more massive than required by stability of the disk component.  This
  suggests a coupling between the formation of disk and bulge, and is
  consistent with the self-regulating, inside-out bulge formation
  scenario proposed here. 

\item If we live in a low-density Universe ($\Omega_0 \lta 0.3$), the
  only efficient way to make spiral galaxies is by assuring that only
  a relatively small fraction of the available baryons makes it into
  the galaxy, and furthermore that the angular momentum distribution
  of those baryons is similar to that of the entire system; i.e., the
  probability that a certain baryon becomes a constituent of the final
  galaxy has to be independent of its specific angular momentum.  In
  the cooling scenario, most of the angular momentum of the system
  remains in the outer layers, and most halos form disk-systems with
  massive bulges, such as S0s.  If, however, the galaxy formation
  efficiency is regulated as described by our `feedback' model, a
  promising scenario unfolds: At formation redshifts $\gtrsim 3.0$ the
  galaxy formation efficiency is unity, and systems that form build up
  a large bulge to support the disk that assembles around them. These
  galaxies resemble S0s.  Galaxies that form later, at $z \approx 0$,
  no longer require a massive bulge, and spirals are preferentially
  formed.  Coupled with the notion that density perturbations that
  collapse early are preferentially found in high density environments
  such as clusters, this scenario automatically predicts a
  morphology-density relation in which S0s are most likely to be found
  in clusters. In a SCDM Universe a similar, albeit less restrictive,
  mechanism is at work, which predicts a morphology-density relation
  of smaller amplitude.

\item A reasonable variation in formation redshift and halo angular
  momentum can yield approximately one order of magnitude
  variation in disk-to-bulge ratio, and our simple formation scenario
  can account for both spirals and S0s. However, disky ellipticals
  have too large bulges and too small disks to be incorporated in this
  scenario.  Apparently, their formation and/or evolution has seen
  some processes that caused the baryons to loose a significant amount
  of their angular momentum.  Merging and galaxy harassment (Moore
  \etal 1996) are likely to play a major role for these systems.
  
\end{itemize}

Finally we wish to emphasize of few of the major shortcomings of the
oversimplified formation scenario discussed here. First of all, we
have neglected the fact that the merging of halos is an ongoing
process, and that this is very effective in destroying disks (e.g.,
Toth \& Ostriker 1992). Gas-dynamical simulations that do not involve
the energy and momentum feedback to the gas from supernova explosions,
stellar winds, UV radiation etc. produce galactic disks that are some
two orders of magnitude smaller than the observed spiral disks (e.g.,
Navarro \& Steinmetz 1997): merging (and also harassment) are very
effective in transporting angular momentum out into the halo, thus
yielding more and more compact galaxies. This problem with forming
galactic disks of proper dimensions is often referred to as the
angular momentum problem.  Despite our ignorance regarding the effects
of merging, harassment, feedback, and (re)-ionization of the Universe,
the observed sizes of (spiral) disks clearly suggest however, that the
combine effect of all these processes is apparently such that the
material that forms galactic disks has not lost much of its original
angular momentum acquired from cosmological torques. The use of the
Fall \& Efstathiou disk formation scenario thus seems justified,
despite the aforementioned shortcomings.

In the past semi-analytical simulations of galaxy formation have been
mainly based on the assumption that all bulges result from the merging
of disk galaxies (e.g., Kauffmann, White \& Guiderdoni 1993; Cole
\etal 1994; Baugh, Cole \& Frenk 1996; Somerville \& Primack 1998).
In this paper we have examined an inside-out bulge formation scenario,
which should be regarded as complimentary to this merging scenario.
Our formation scenario can account for both spirals and S0s,
but fails to build systems that are even more bulge dominated.
Although it has been shown that bar-instabilities can lead to the
formation of small (and close to exponential) bulges, a more detailed
study is required to investigate whether the inside-out bulge
formation scenario discussed here can yield more massive bulges as
well. It is at least intriguing that the critical spin parameter is of
the same order of magnitude as the typical spin parameter for halos,
suggesting that a significant fraction of halos will yield unstable
disks, unless part of the baryonic material is transformed into a bulge
component as suggested here. So despite the clearly oversimplified
nature of the formation scenario envisioned here, it may provide a
useful framework for future investigations of galaxy formation.


\section{Acknowledgments}

For discussions and advice, I'm indebted to George Lake, Vincent Icke,
Stephan Courteau, Julio Navarro, and to the referee. This work was
supported by a Hubble Fellowship, {\#}HF-01102.11-97A, awarded by
STScI.
 

\clearpage

\appendix

\section{Angular momentum distribution of the disk}

Mestel (1963) noticed that disks of spirals have specific angular
momentum distributions similar to that of a uniform sphere in solid
body rotation.  Exponential disks are thus a natural consequence of
the Fall \& Efstathiou disk formation scenario, if we assume that
proto-galaxies have a specific angular momentum distribution similar
to that of a uniformly rotating, uniform sphere, and that the material
that forms the disk does not loose its specific angular momentum
(e.g., Gunn 1982; Dalcanton \etal 1997).

The total angular momentum of a uniformly rotating, uniform sphere
with radius $a$ and constant angular velocity $\omega$ is
\begin{equation}
\label{totangmom}
J =  {2\over 5} M \omega a^2,
\end{equation}
and the mass with specific angular momentum less than $j$ is given by
\begin{equation}
\label{angmomdistr}
M(< j) = M \Biggl[ 1 - \biggl(1 - {j\over \omega a^2}\biggr)^{2/3}\Biggr].
\end{equation}
Since the baryonic and dark matter are well mixed, the angular
momentum distribution for the baryonic matter component is identical
to that of the total mass of the system, and $J_{\rm bar} =
(\Omega_{\rm bar}/\Omega_0) J$.  The bulge forms out of the low angular
momentum material with a {\it cylindrical} radius $R < R_{\rm
  bulge}$.  Solving $M_{\rm bar}(< j) = M_b$ yields
\begin{equation}
\label{bulgerad}
R_{\rm bulge} = a \sqrt{1 - \biggl({t \over 1 + t}\biggr)^{3/2}},
\end{equation}
and the total angular momentum of the bulge is
\begin{equation}
\label{angmombulge}
J_b = {2\over 5} M_{\rm bar} a^2 \omega \Biggl[ 1 - \biggl(
{3\over 2}\Bigl({R_{\rm bulge} \over a}\Bigr)^2 + 1\biggr) \biggl( 1
- \Bigl({R_{\rm bulge} \over a}\Bigr)^2\biggr)^{3/2}\Biggr],
\end{equation}
which is the total angular momentum of the intersection of a solid
sphere with radius $a$ and a solid cylinder with radius $R_{\rm bulge}
\leq a$ (where the sphere and cylinder have a common center).  The
disk forms out of the material with $R_{\rm bulge} < R < a$, and has a
total angular momentum $J_d = J_{\rm bar} - J_b \equiv j_d J$. We
assume that the specific angular momentum of this material is
conserved during disk formation.

In the feedback scenario, in which each baryon has a probability
$\epsilon_{gf}$ of ending up in the final disk-bulge system, we find
upon substituting $R_{\rm bulge}$ (equation [\ref{bulgerad}]) in
equation (\ref{angmombulge}), and with the use of
equations~(\ref{dismass}) and~(\ref{bulmass})
\begin{equation}
\label{amomdisk}
j_d = [m_d + m_b] \; {t\over 1 + t} \; \Biggl[ {5\over 2} - {3 \over 2}
\biggl({t \over 1 + t}\biggr)^{2/3}\Biggr] \equiv f(t,\epsilon_{gf}) \;
[m_d + M_b]
\end{equation}

In the cooling scenario, only the baryonic material with $r < r_{\rm
  cool}$ cools to form the disk and bulge, where
\begin{equation}
\label{coolrad}
r_{\rm cool} = \epsilon_{gf}^{1/3} a,
\end{equation}
and one finds
\begin{equation}
\label{amomdisk2}
j_d = \epsilon_{gf}^{2/3} (m_d + m_b) {t\over 1 + t} \Biggl[ {5\over
  2} - {3 \over 2} \biggl({t \over 1 + t}\biggr)^{2/3}\Biggr] \equiv
f(t,\epsilon_{gf}) [m_d + M_b].
\end{equation}

\clearpage


\ifsubmode\else
\baselineskip=10pt
\fi


\clearpage


\ifsubmode\else
\baselineskip=14pt
\fi


\figcaption[vdb98f1]{The critical value $\lambda_{\rm crit}$ of the
  halo's spin parameter (plotted logarithmically) as a function of
  redshift $z$ for both the feedback scenario (left panel) and the
  cooling scenario (right panel). Results are plotted for three
  different cosmologies: SCDM (solid line), OCDM (dot-dashed line),
  and $\Lambda$CDM (dashed line). The five dotted lines are labeled by
  the probability of finding a halo below that line for the spin
  parameter distribution discussed in Section~2.2. The increase of
  $\lambda_{\rm crit}$ with increasing redshift is due to the
  increasing galaxy formation efficiency. When this efficiency becomes
  unity, $\lambda_{\rm crit}$ becomes constant with redshift as
  apparent in the figure. Note the similarity between the OCDM and
  $\Lambda$CDM cosmologies compared to the SCDM scenario.
  \label{fig1}}

\figcaption[vdb98f2]{The $\mu_0$--$R_d$ diagram for a large variety
  of disks: LSB spirals (triangles), HSB spirals (open circles), S0s
  (stars), disky ellipticals (solid circles), and nuclear disks
  (squares).  The dotted line (partly plotted as a thick solid line)
  corresponds to the line with $L_d = 3 L^{*}$. The dashed line
  corresponds to $\lambda = \lambda_{\rm crit}$ for the case without
  bulge (i.e., $t=\infty$). The curved thick solid line corresponds to
  $\lambda = \lambda_{\rm crit}$ and $L_d + L_b = 3 L^{*}$, with $t$
  running from zero to infinity. These lines border 4 regions, the
  meaning of which are discussed in the text.\label{fig2}}

\figcaption[vdb98f3]{Results for a SCDM cosmology. Plotted are the
  logarithm of the spin parameter versus the logarithm of the
  disk-to-bulge ratio ($D/B = t$). Solid circles correspond to disky
  ellipticals, stars to S0s, open circles to HSB spirals, and
  triangles to LSB spirals.  The thick solid line is the stability
  margin $\lambda_{\rm crit}$; halos below this line result in
  unstable disks.  As can be seen, real disks avoid this region, but
  stay relatively close to the stability margin, in agreement with the
  self-regulating bulge formation scenario discussed in Section~2.1.
  The dashed curves correspond to the 1, 10, 50, 90, and 99 percent
  levels of the cumulative distribution of the spin parameter
  according to equation (\ref{spindistr}) with $\bar{\lambda} = 0.05$
  and $\sigma_{\lambda} = 0.7$. Upper panels have a galaxy formation
  efficiency equal to unity, middle panels correspond to the cooling
  scenario, and lower panels to the feedback scenario. Panels on the
  left correspond to $z_{\rm form} = 0$, middle panels to $z_{\rm
    form} = 1$, and panels on the right to $z_{\rm form} = 3$.
  \label{fig3}}

\figcaption[vdb98f4]{Same as Figure~3, except now for the OCDM universe
  with $\Omega_0 = 0.3$. The main difference is that $\lambda_{\rm
    crit}$ reaches higher values here. Both in the case with
  $\epsilon_{gf} = 1.0$ (upper panels), and in the cooling scenario
  (middle panels), $\sim 90$ percent of the halos is expected to
  result in unstable disks unless a significant bulge is formed (in
  the cooling scenario this reduces to $\sim 75$ percent for $z_{\rm
    form} = 0$).  The majority of the halos yields systems with a
  significant bulge such as S0s, and both these models can thus be
  ruled out, given the relatively large numbers of spirals in the
  present Universe. In the feedback scenario (lower panels) stable
  spiral systems {\it can} form, albeit only at low redshifts.  At
  higher redshifts systems with smaller disk-to-bulge ratios form,
  such as S0s.
  \label{fig4}}

\figcaption[vdb98f5]{Histograms of the (normalized) distribution of
  formation redshifts (plotted as $\log\bigl[1 + z_{\rm form}\bigr]$)
  determined as described in the text for spirals (left panel, HSB and
  LSB spirals combined), S0s (middle panel), and disky ellipticals
  (right panel).  The arrows indicate the means for each sample (the
  mean for the spirals is zero).  The number $n$ of galaxies in each
  of the separate samples is indicated in the corresponding boxes. The
  dotted lines indicate formation redshifts of $z=1$, $z=3$, and
  $z=10$. These histograms correspond to the SCDM model with feedback,
  but are similar for other cosmogonies.
  \label{fig5}}

\figcaption[vdb98f6]{The specific angular momentum $J/M$ (in kpc km
  s$^{-1}$) of the disk versus that of the bulge. As can be seen,
  within the errorbars the specific angular momenta of disk and bulge
  are roughly similar. (Data taken from Scorza \& Bender 1995).
  \label{fig6}}


\ifprintfig

\clearpage

\plotone{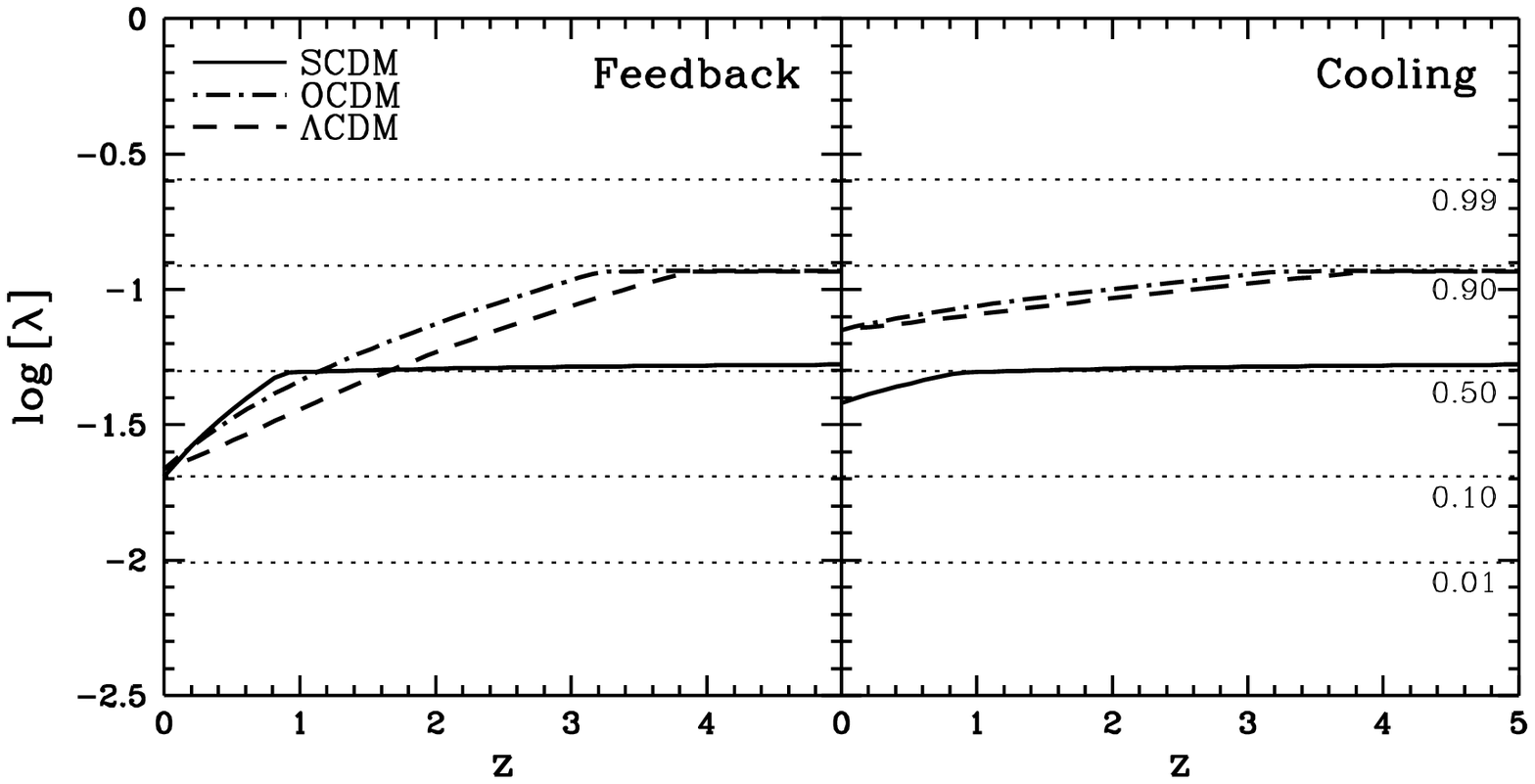}

\clearpage

\plotone{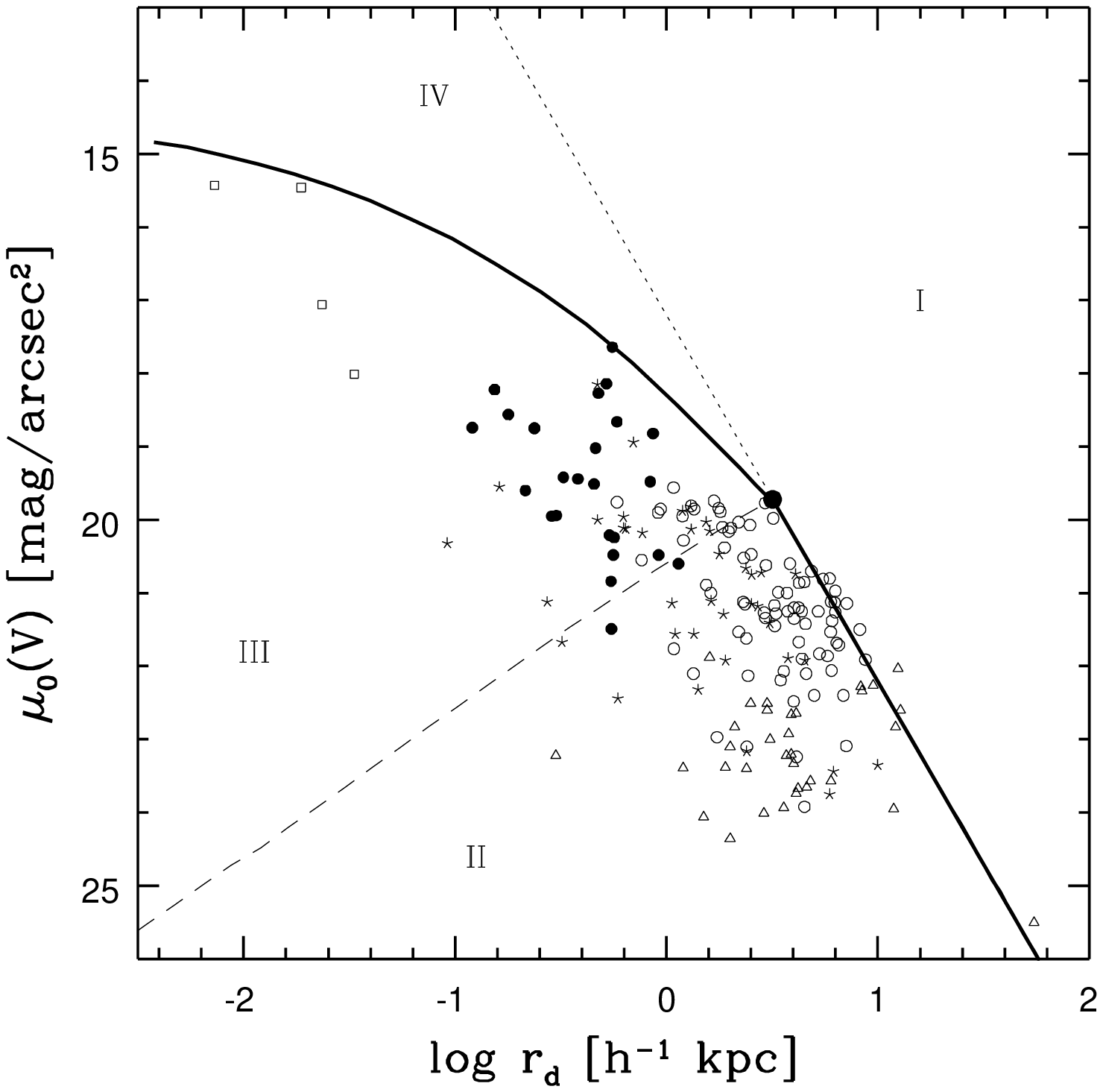}

\clearpage

\plotone{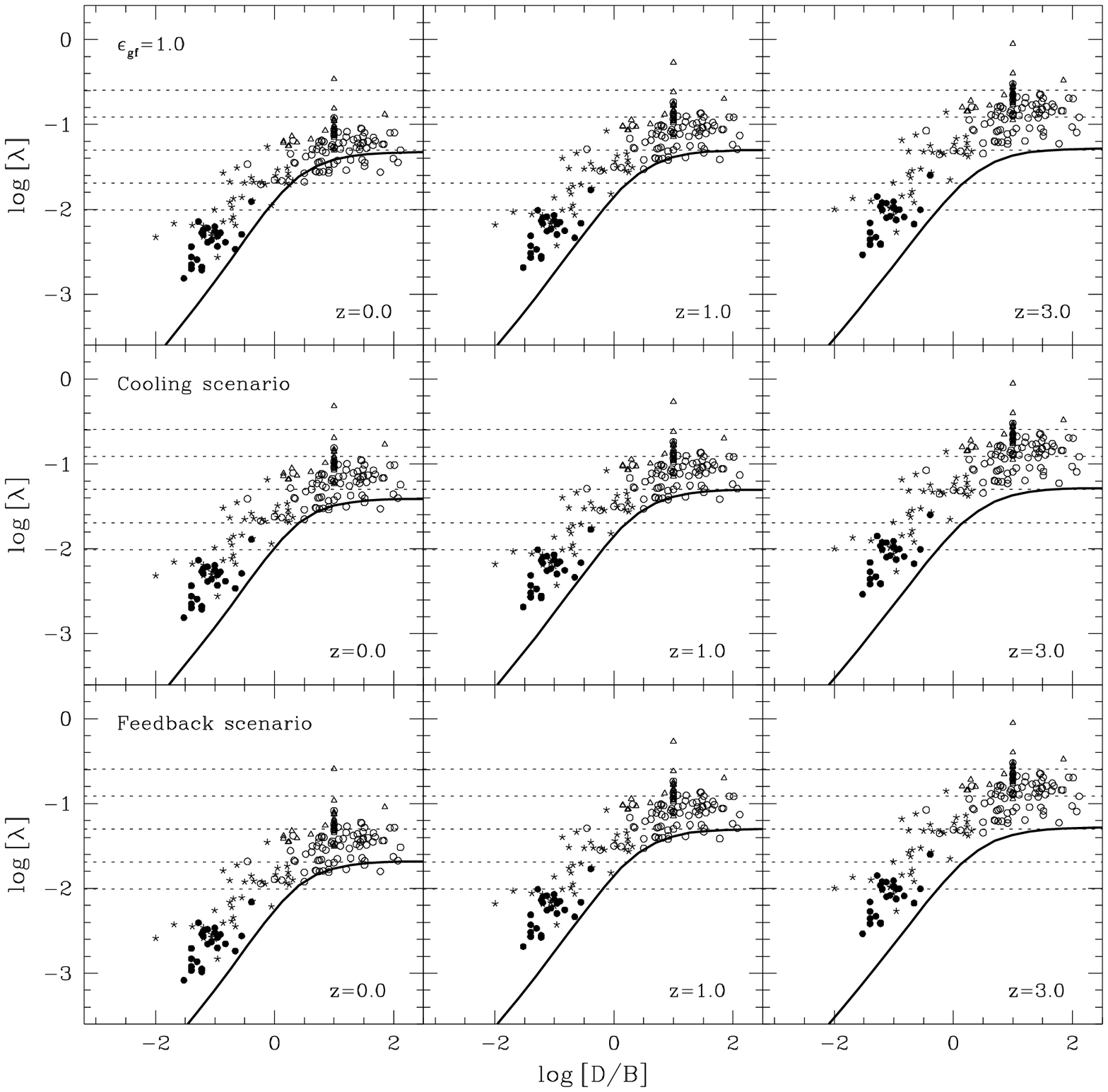}

\clearpage

\plotone{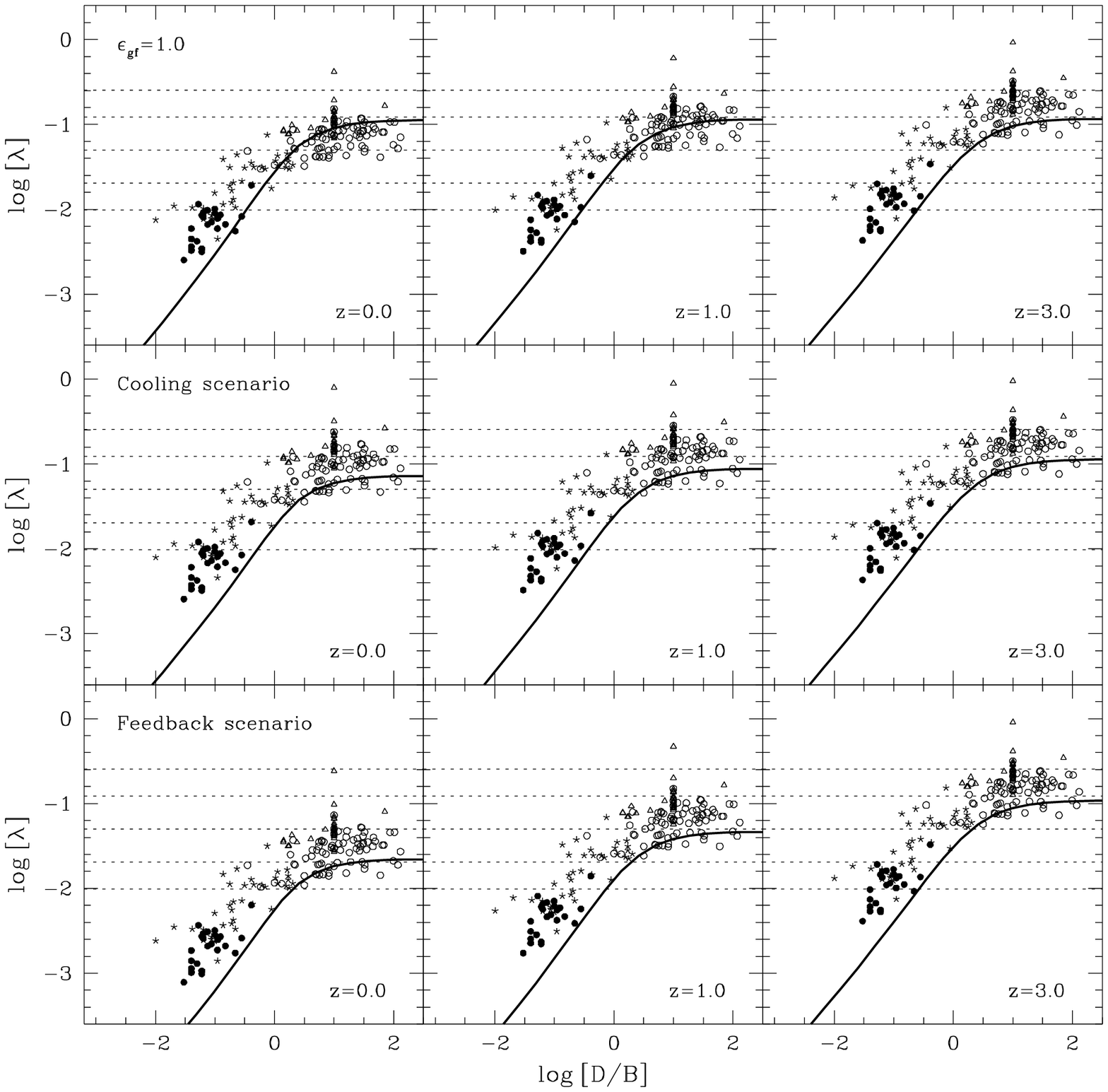}

\clearpage

\plotone{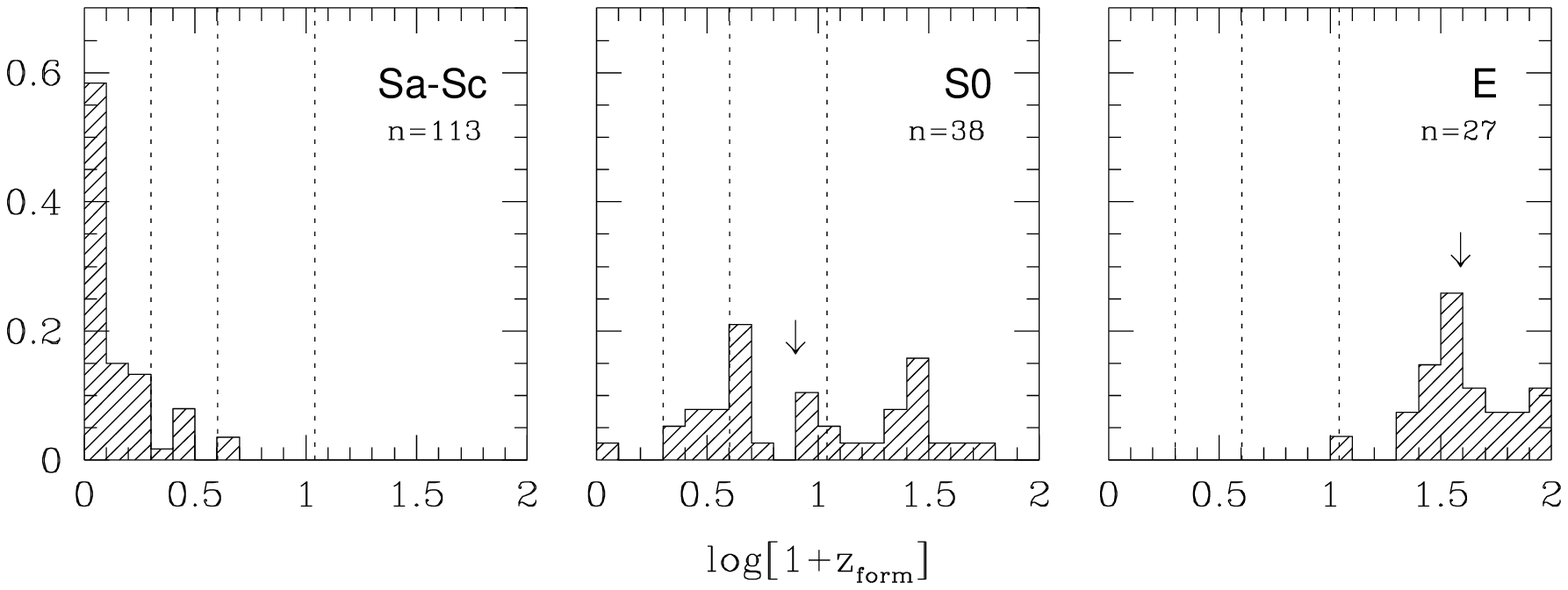}

\clearpage

\plotone{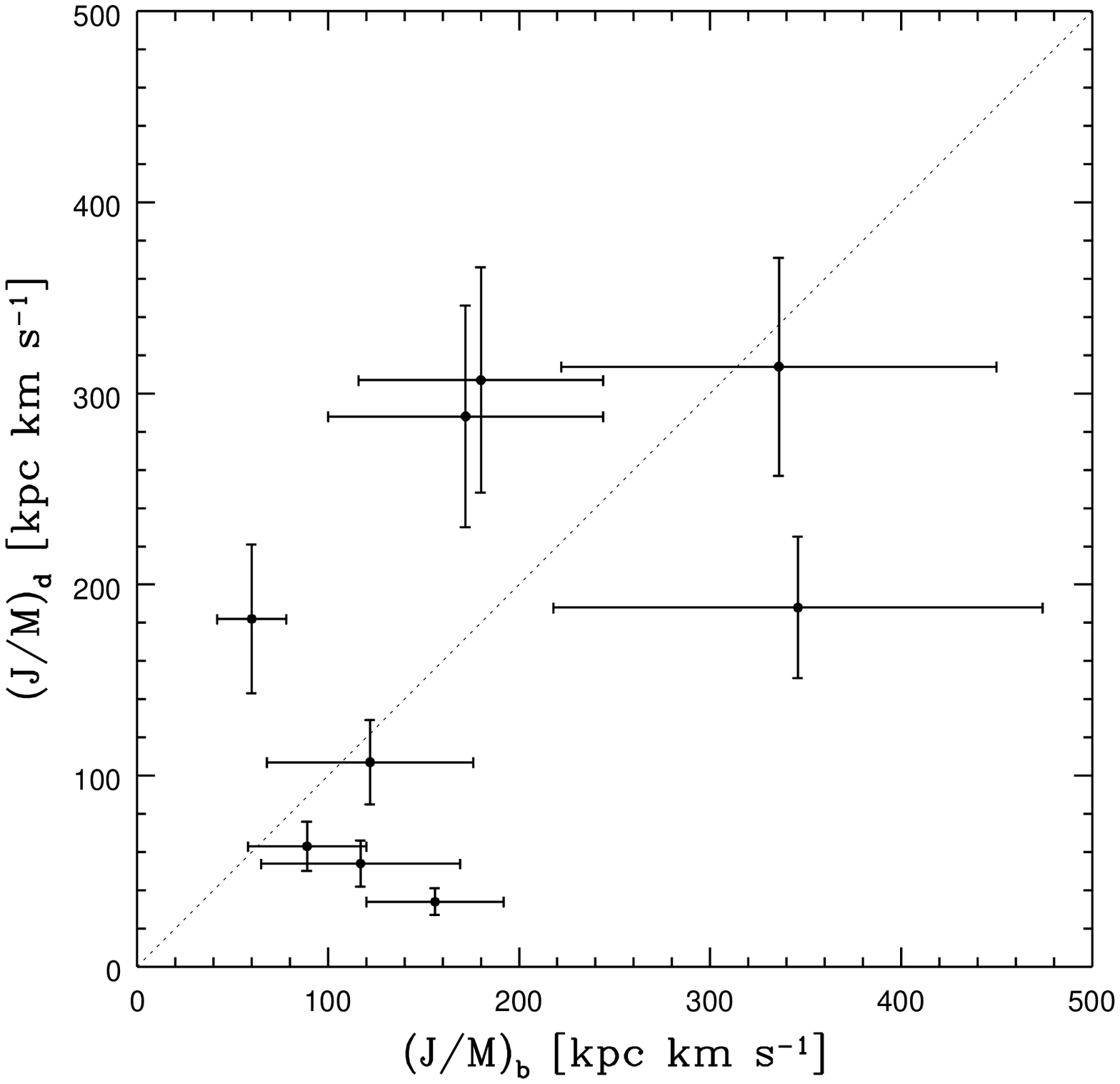}

\fi


\end{document}